\documentclass[11pt]{article}

\usepackage{fullpage,amsmath,xspace,amssymb,epsfig}

\newtheorem{theorem}{Theorem}
\newtheorem{definition}{Definition}
\newtheorem{corollary}[theorem]{Corollary}    
\newcommand{\qed}{\hfill{$\rule{6pt}{6pt}$}} 
\newenvironment{proof}{\noindent{\bf Proof:}}{\qed\\}

\newcommand\tribes{\mbox{\sf Tribes}\xspace}

\newcommand{\suppress}[1]{}
\newcommand{\comment}[1]{}

\newcommand{\cX}{{{\mathcal X}}}
\newcommand{\sD}{{{\mathsf D}}}
\newcommand{\sR}{{{\mathsf R}}}

\newcommand{\eps}{\varepsilon}
\newcommand{\cZ}{\mathcal{Z}}
\newcommand{\cP}{\mathcal{P}}
\newcommand{\ans}{\mathrm{ans}}
\newcommand\pr{\mbox{Pr}}
\newcommand\var{\mbox{Var}}

\begin{document}
\title{{\bf The influence lower bound via query elimination}\thanks{The work was done when R.J. visited The Chinese University of Hong Kong and S.Z. visited Centre of Quantum Technologies (CQT) under support of CQT and Hong Kong General Research Fund 419309 and 418710.}
}

\author{Rahul Jain\thanks{Centre for Quantum Technologies and Department of Computer Science, National University of Singapore, Singapore. Email: {\tt rahul@comp.nus.edu.sg}.}
\and
Shengyu Zhang\thanks{Department of Computer Science and Engineering and The Institute of Theoretical Computer Science and Communications, The Chinese University of Hong Kong, Shatin, Hong Kong. Email: {\tt syzhang@cse.cuhk.edu.hk}. }
}

\date{}

\maketitle

\begin{abstract}
We give a simpler proof, via query elimination, of a result due to O'Donnell, Saks, Schramm and Servedio, which shows a lower bound on the zero-error randomized query complexity of a function $f$ in terms of the maximum influence of any variable of $f$. Our lower bound also applies to the two-sided error distributional query complexity of $f$, and it allows an immediate extension which can be used to prove stronger lower bounds for some functions. 

\end{abstract}

\section{Introduction}\label{sec: intro}
Query complexity measures the hardness of computing a function $f$ by the minimum number of input variables one needs to read before knowing the function value. A $k$-query $\epsilon$-error randomized query algorithm is one that, on all inputs, has at most $\epsilon$ error probability and makes at most $k$ queries over all random coins. The $\epsilon$-error randomized query complexity of $f$, denoted $\sR_\eps(f)$, is the minimum number $k$ such that there exists a $k$-query $\epsilon$-error randomized query algorithm. The influence of a variable is another important quantity which measures the importance of the variable to the function value (on average over other variables). More precisely, for a function $f: \cX^n\rightarrow \cZ$ and a distribution $\mu$ on $\cX$, the influence of the $i$-th variable is defined as $\inf_i(f,\mu) = \pr[f(X) \neq f(X^i)]$, where $X = X_1 \ldots X_n$ is drawn from $\mu^{\otimes n}$ and $X^i$ is obtained from $X$ by re-randomizing $X_i$; namely $X^i = X_1 \ldots X_{i-1}Y_iX_{i+1} \ldots X_n$, where $Y_i$ is drawn from $\mu$ and $Y_i$ is independent of $X$. Both query complexity and influence are well-studied subjects; see \cite{BdW02} for a survey of the former (with many other complexity measures) and \cite{ODo08} for a survey of the latter (and Fourier analysis on Boolean functions).

Randomized query complexity can be lower bounded in terms of influence. In \cite{OSSS05}, O'Donnell, Saks, Schramm and Servedio proved that for all Boolean functions $f:\{-1,+1\}^n \rightarrow \{-1,+1\}$, 
\begin{equation}
	\sR_0(f) = \Omega\left(\frac{\var[f]}{\max_i \inf_i(f,\mu_p)}\right) \enspace .
\label{eq:lb1}
\end{equation} 
Above $\mu_p$ is the distribution on $\{-1,+1\}$ with $-1$ picked with probability $p$; $\var[f]$ is the variance of $f(X)$ with $X$ drawn from $\mu_p^{\otimes n}$ and $\sR_0(f)$ represents the zero-error randomized query complexity of $f$; namely the minimum over all randomized query algorithms with no error on each input, of the  maximum expected (over the random coins) number of queries made by the query algorithm on any input. 
Recently Lee \cite{Lee10} gave another proof of this fact. Together with another bound $\sR_0(f) \geq (\sum_i \inf_i(f,\mu_p))^2/(4p(1-p))$ for monotone functions \cite{OS07}, it gives a lower bound of $\Omega(n^{2/3})$ for all monotone functions invariant to a transitive group of permutations (on variables). This in particular reproduces the $\Omega(n^{4/3})$ lower bound for all monotone graph properties in \cite{Haj91}, which is $O(\log^{1/3}(n))$ shy of record \cite{CK01}.

In this paper we give a new proof of Eq. \eqref{eq:lb1}, arguably shorter and simpler than both previous ones \cite{OSSS05,Lee10}. In fact we prove a stronger statement that applies to the two-sided error case. The basic idea is by query elimination: we can save one query without increasing the error by more than $\max_i \inf_i(f,p)$, and eventually eliminate all queries to obtain a zero-query algorithm, which must have a large error probability on a hard distribution. This lower bounds the number of queries of the original algorithm. The analysis for the increase in error due to eliminating one query is quite simple and follows from the union bound (applied just once) and the observation that $X^i$ is identically distributed to $X$. 

Since we lower bound distributional query complexity (defined in the next section), we get a smoothed version of the influence bound as an immediate consequence. As in the cases with the rectangle bound and the discrepancy bound in communication complexity and query complexity, where the smoothed versions can prove strong lower bounds~\cite{Klauck07, Sherstov08, SZ09, LZ10, Klauck10, JainK10, CR11}, this smoothed influence lower bound also gives stronger bounds for some functions than Eq. \eqref{eq:lb1}. 

\section{Main result and proof} 

\begin{definition}[Influence]
Let $f : \cX^n \rightarrow \cZ$ be a function, and $X_i$'s and $Y_i$'s (for $i= 1, ..., n$) be random variables i.i.d. distributed according to $\mu$ on $\cX$. For each $i \in [n]$, let $X^i$ represent the random variable $X_1 \ldots X_{i-1} Y_i X_{i+1} \ldots X_n$. The {\em influence} of variable $X_i$ on $f$ is defined as $\inf_i(f,\mu) = \Pr[f(X) \neq f(X^i)]$. The {\em maximum influence} of $f$ with respect to $\mu$ is defined as $\inf_{\max}(f,\mu) = \max_i\inf_i(f,\mu)$.
\end{definition}

For $\eps > 0 $, a deterministic $k$-query algorithm has the $\lambda$-distributional error $\epsilon$ if it makes at most $k$ queries over all possible inputs, and for a random input drawn from $\lambda$, the average error probability is $\epsilon$.  The $\epsilon$-error $\lambda$-distributional query complexity of $f$, denoted $\sD^\lambda_\eps(f)$, is the minimum number $k$ such that there exists a $k$-query algorithm which  has the $\mu$-distributional error $\epsilon$. We show the following.

\begin{theorem} Let $f : \cX^n \rightarrow \cZ$ be a function, $\mu$ be a distribution on $\cX$ and $\eps > 0$. Let $X$ be drawn from $\mu^{\otimes n}$.  Then,
\[\sD^{\mu^{\otimes n}}_\eps(f) \geq \frac{1 - \max_{z\in \cZ}\Pr[f(X)=z] - \eps}{\inf_{\max}(f,\mu)}.\]
\label{thm:main}
\end{theorem}
\begin{proof}
Let $\cP_k$ be a deterministic $k$-query algorithm for $f$ with $\mu^{\otimes n}$-distributional error at most $\delta$. We present a deterministic $(k-1)$-query algorithm $\cP_{k-1}$ for $f$ with $\mu^{\otimes n}$-distributional error at most $\delta + \inf_{\max}(f,\mu)$. This way, starting from an algorithm which makes $\sD^{\mu^{\otimes n}}_\eps(f)$ queries and has average error at most $\eps$, repeating the above procedure gives another algorithm $\cP_0 $ which makes no queries and has average error at most $\eps + \sD^{\mu^{\otimes n}}_\eps(f) \cdot  \inf_{\max}(f,\mu)$. It is easily seen that  $\cP_0$ must have error at least $1 - \max_{z\in \cZ}\Pr[f(X)=z] $ and hence we get the desired result. 

Now we show how to obtain $\cP_{k-1}$ from $\cP_k$. We will show a randomized algorithm $\cP_{k-1}'$ with at most $k-1$ queries on any input and any random coins and average error under $\mu^{\otimes n}$  at most $\delta + \inf_{\max}(f,\mu)$. From $\cP_{k-1}'$, using an easy averaging argument (and fixing coins of $\cP_{k-1}'$ appropriately), we can get a deterministic algorithm $\cP_{k-1}$ with at most $k-1$ queries on any input and the same average error bound as in $\cP_{k-1}'$. 

Let $X_i$ be the first query of $\cP_k$ and without loss of generality we can assume that $\cP_k$ does not query $X_i$ any more afterward. In $\cP_{k-1}'$ we do not make this query, but assume the answer to this query to be $Y_i$, where $Y_i$ is distributed according to $\mu$ and is independent of $X$. From here on $\cP_{k-1}'$ proceeds identically to $\cP_k$. 
By construction the maximum number of queries made by $\cP_{k-1}'$ is at most $k-1$. Let $\ans(\cP, X)$ represent the answer of algorithm $\cP$ on input $X$. Since $\ans(\cP_k, X^i) \neq f(X)$ implies either $\ans(\cP_k, X^i) \neq f(X^i)$ or $f(X^i) \neq f(X)$, we have
\begin{align*}
\lefteqn{\Pr[\cP_{k-1}' \text{ makes error on input } X] = ~ \Pr[\ans(\cP_k, X^i) \neq f(X)] } \\
\leq & ~\Pr[ \ans(\cP_k, X^i) \neq f(X^i)] + \Pr[ f(X^i) \neq f(X) ] \quad \mbox{(from union bound)} \\
= & ~\Pr[\ans(\cP_k, X) \neq f(X)] + \Pr[f(X^i) \neq f(X) ] \quad \mbox{(since $X$ is identically distributed to $X^i$)} \\
\leq & ~ \delta + \inf{_{\max}}(f,\mu) \enspace .
\end{align*}
\end{proof}
It is easily argued that $\sR_0(f) = \Omega(\sR_\eps(f)) = \Omega(\sD^{\mu^{\otimes n}}_\eps(f))$ for $\eps, \mu$ as above. Also $1-\max_z\Pr[f(X)=z] = \Omega(\var[f])$ for Boolean functions $f$, therefore the above theorem implies Eq. \eqref{eq:lb1}.

\medskip
Next we improve the lower bound by going to a function $g$, which is close to $f$ but could potentially have smaller $\inf_{\max}$. 
Let $g: \cX^{n} \rightarrow \cZ$ be a function such that $\Pr[f(X) \neq g(X)] \leq \delta$, where $X$ is drawn from $\mu^{\otimes n}$ as above and $\delta \geq 0$.  It is easily noted that an algorithm for $f$ with average error under $\mu^{\otimes n}$ being at most $\eps$ also works as an algorithm for $g$ with average error under $\mu^{\otimes n}$ being at most $\eps + \delta$. Therefore $\sD^{\mu^{\otimes n}}_\eps(f) \geq \sD^{\mu^{\otimes n}}_{\eps + \delta} (g)  $. Hence as a corollary of Theorem~\ref{thm:main} we get that a smoothed version of the influence bound also applies as a lower bound on the distributional query complexity of $f$.
\begin{corollary}
Let $f : \cX^n \rightarrow \cZ$ be a function, $\mu$ be a distribution on $\cX$ and $\eps > 0, \delta \geq 0$. Let $X$ be drawn from $\mu^{\otimes n}$. Let $g: \cX^{n} \rightarrow \cZ$ be a function such that $\Pr[f(X) \neq g(X)] \leq \delta$.  Then
\[\sD^{\mu^{\otimes n}}_\eps(f) \geq \sD^{\mu^{\otimes n}}_{\eps + \delta} (g)  \geq \frac{1 - \max_{z\in \cZ}\Pr[g(X)=z] - \eps - \delta}{\inf_{\max}(g,\mu)}.\]
\end{corollary}

Note that there are functions $f$ with large $\inf_{\max}$ but close to some other function $g$ with small $\inf_{\max}$. For example, \tribes is OR of $s\approx n/\log_2 n$ AND gates, each of degree $t \approx \log_2 n - \log_2\log_2n$. The parameters $s,t$ are so set to make exactly half the inputs being 1. It is well known that for this function, all influences $\inf_i = \Theta(\log n/n)$, where the distribution is uniform on all inputs. Let $g$ be \tribes, and obtain $f$ from $g$ by picking a $\delta$-fraction of inputs $x$ and changing their function values to $f(x) = x_1$ ($x_1$ is the first bit of $x$) and for the rest $f(x) = g(x)$. Then the first variable has influence at least $\Omega(\delta)$, so applying the old bound only gives a constant lower bound. But $g$ is $\delta$-close to $f$ with $\inf_{\max}(g) = \Theta(\log n/n)$. So the above corollary gives a much better lower bound of $\Theta(n/\log n)$ for the distributional query complexity of $f$, which the original bound Eq. \eqref{eq:lb1} only gives a constant. 

A final comment is that our proof does not assume that the distributions of the different variables are the same. The proof goes through and the bound applies analogously as long as these distributions are independent.

\medskip

\noindent {\bf Acknowledgment: } We thank Ronald de Wolf for detailed and helpful comments on an earlier draft of the paper.
\bibliography{inf}
\bibliographystyle{plain}

\end{document}